\documentclass[amsmath, amsthm, amssymb, floatfix, reprint, twocolumn, notitlepage, nofootinbib, 10pt]{revtex4-1}

\newtheorem{theorem}{Theorem}
\newtheorem{lemma}{Lemma}

\usepackage[utf8]{inputenc}
\usepackage{microtype,bm,bbm,graphicx,booktabs,times}
\usepackage[usenames,dvipsnames]{xcolor}
\usepackage{setspace}
\selectfont{}

\usepackage{hyperref}
\usepackage{braket}
\usepackage{subfiles} 
\usepackage{upgreek}
\usepackage{color}
\hypersetup{colorlinks,allcolors=black,linktoc=all}
\usepackage[capitalize]{cleveref}
\crefformat{section}{Sec.~#2#1#3}

\newcommand{\edit}[1]{{\color{black} #1}}

\newcommand{\norm}[1]{\left\Vert{#1}\right\Vert}

\begin{document}
\title{Quantum algorithms for powering stable Hermitian matrices}
\author{Guillermo González$^{1, 2, \dagger}$}
\author{Rahul Trivedi$^{1, 2, \dagger}$}
\email{rahul.trivedi@mpq.mpg.de}
\author{J.~Ignacio Cirac$^{1, 2}$}
\address{$^1$Max-Planck-Institut für Quantenoptik, Hans-Kopfermann-Str.~1, 85748 Garching, Germany.\\
$^2$Munich Center for Quantum Science and Technology (MCQST), Schellingstr. 4, D-80799 Munich, Germany.\\
$^\dagger$equal contribution}

\date{\today}

\begin{abstract}
Matrix powering is a fundamental computational primitive in linear algebra. It has widespread applications in scientific computing and engineering, and underlies the solution of time-homogeneous linear ordinary differential equations, simulation of discrete-time Markov chains, or discovering the spectral properties of matrices with iterative methods. In this paper, we investigate the possibility of speeding up matrix powering of sparse stable Hermitian matrices on a quantum computer. We present two quantum algorithms that can achieve speedup over the classical matrix powering algorithms --- (i) an adaption of quantum-walk based fast forwarding algorithm (ii) an algorithm based on Hamiltonian simulation. Furthermore, by mapping the $N-$bit parity determination problem to a matrix powering problem, we provide no-go theorems that limit the quantum speedups achievable in powering non-Hermitian matrices.
\end{abstract}
\maketitle

\section{Introduction}
Recent years have seen rapid progress in the development of quantum computing hardware, and there have already been experimental demonstrations of quantum computations that are believed to be hard to simulate on classical computers \cite{google2019quantumsupremacy,Zhong2020quantumsupremacy}. While this progress in hardware has brought us closer to the monumental goal of building a fault tolerant quantum computer, it has also provided us with access to noisy quantum hardware which might already solve problems that are hard for classical computers \cite{Preskill2018NISQ}. From a theoretical standpoint, it has become important to discover algorithms that can provide speedup over their classical counterparts on both fault tolerant quantum computers and near-term noisy quantum hardware.

Quantum computers are known to offer exponential speedup in simulating the physics of quantum systems --- near-optimal algorithms have been developed for the simulation of Hamiltonian dynamics \cite{Lloyd1996simulators,Berry2015hamiltonian,Childs2012HamiltonianLCU}, Lindbladian dynamics \cite{kliesch2011quantum-church-turing,di2015quantum, chenu2017quantum, cleve2016efficient} and steady state (finite temperature or ground state) properties of Hamiltonians \cite{ge2019faster, oh2008quantum, schutzhold2006adiabatic, lu2020algorithms}. Several techniques used for simulating quantum systems have been generalized to accelerate more fundamental linear algebra computational primitives --- exponential quantum speedup in the solution of systems of linear equations  have been obtained \cite{Harrow2009linear-systems, childs2017qlsp, ambainis2012variable}, and quantum speedups have also been shown in solving ordinary differential equations \cite{Berry2014ODE, berry2017quantum} and partial differential equations \cite{childs2020high}.

Another fundamental computation that can be accelerated on quantum computers is matrix powering i.e.~computing a matrix-element $v^\dagger A^t u$ given access to the matrix $A \in \mathbb{C}^{N\times N}$, a positive integer power $t$ and vectors $v, u \in \mathbb{C}^N$. This is a computational primitive which appears in various applications, including but not limited to solving linear differential equations, simulating discrete-time Markov chains as well as matrix inversion and eigenvalue computation using Krylov subspace methods. Classically, this problem can be solved by repeated matrix multiplication in time $O(\text{poly}(N)Dt)$, where $D$ is the sparsity of the matrix $A$. Without any assumptions on the matrix $A$, one approach to solve the matrix-powering problem on a quantum computer is to map it to a matrix inversion problem and use quantum algorithms for solving linear equations --- this approach has been investigated in Refs.~\cite{Berry2014ODE, berry2017quantum} in the context of solving linear time-homogeneous ODEs and has a run-time $O(\text{polylog}(N)\text{poly}(\varepsilon^{-1})\kappa_V\norm{v}^2\norm{u}^2 Dt)$, where $\kappa_V$ is the condition number of the eigenvector matrix of $A$, to obtain $v^\dagger A^t u$ to a precision $\varepsilon$ for stable matrices thereby providing an exponential speedup in the matrix size over classical algorithms.

Furthermore, several authors have studied the problem of powering a stochastic matrix which arises in the context of simulating the dynamics of a discrete-time Markov chain \cite{van1992stochastic,risken1996fokker, gagniuc2017markov}. Classically, the problem of powering a stochastic matrix can be solved efficiently to precision $\varepsilon$ with the Monte Carlo algorithm in time $O(Dt \norm{v}^2\norm{u}^2 / \varepsilon^2)$ --- using the Monte Carlo algorithm is thus exponentially faster than using repeated matrix multiplication. While the quantum algorithms based on linear-equation solve do not provide an exponential speedup for stochastic matrix powering when compared to the classical Monte Carlo algorithm, there have been two proposals for achieving polynomial quantum speedups for this specific problem. One of the proposed algorithms is to use a reversible implementation of the classical Monte Carlo together with quantum amplitude estimation to achieve a quadratic improvement in the dependence of the run-time on precision as compared to the classical Monte Carlo algorithm \cite{Montanaro2015Montecarlo}. This idea has been applied to propose solutions to the heat equation \cite{linden2020heat-eq}, and stochastic differential equations \cite{an2020quantum-accelerated-montecarlo}. A different quantum speedup can be obtained for symmetric stochastic matrices by employing quantum walks \cite{watrous2001qwalk, szegedy2004qwalk, subramanian2019implementing, aharonov2001qwalks-on-graphs}. In particular, Ref.~\cite{apers2018quantum} prepares a quantum state within an $\varepsilon-$radius of \edit{$A^t u / \norm{A^t u}$ in $O(D \norm{A^{t}u} ^{-1}\sqrt{t}\ \text{log}^{1/2}(\varepsilon^{-1}\norm{A^tu}^{-1}))$}. The same author generalized this algorithm to arbitrary Hermitian matrices $A$ in Ref.~\cite{apers2019quantum}. While this algorithm obtains a quantum speedup over classical methods (an exponential speedup in $N$ over repeated matrix multiplication, and quadratic speedup in $t$ over Monte Carlo algorithm), the dependence of the run-time on $\norm{A^t u}^{-1}$ can often make it polynomially slow in $N$.

In this paper we introduce two algorithms to compute $v^\dagger A^t u$ for stable Hermitian matrices $A$ i.e.~Hermitian matrices all of whose eigenvalues have magnitudes less than 1. The first algorithm, which combines the construction of Ref.~\cite{apers2018quantum} with a Hadamard test \cite{aharonov2009polynomial}, has a run-time $\widetilde{O}(D\sqrt{t}\norm{v} \norm{u} \varepsilon^{-1} \norm{A}_1^t)$\footnote{\emph{Notation for norms}: Throughout this paper, for a vector $v \in \mathbb{C}^N$, $\norm{v}_k, k \in \{1, 2 \dots\}$ will refer to the standard $\ell^k$ norm of the vector. Furthermore, for convenience, we will use $\norm{v}$ to denote the $\ell^2$ norm of $v$. For matrices $A \in \mathbb{C}^{N\times N}$, $\norm{A}_k$ denotes the operator norm induced by $\ell^k$ vector norm i.e.~$\norm{A}_k = \sup_{v} \norm{Av}_k / \norm{v}_k$. In particular, $\norm{A}_2$ will be the largest singular value of $A$, which coincides with the largest magnitude eigenvalue of $A$ is Hermitian. Additionally, for Hermitian matrices $A$, $\norm{A}_\infty = \norm{A}_1$ will be the maximum absolute row (or column) sum of the matrix $A$.}. For matrices where it is known that $\norm{A}_1 \leq 1$, this provides a quantum speedup over repeated matrix multiplication, since its run-time does not scale polynomially with the size of the matrix. Furthermore, it provides a quadratic speedup in $t$ over classical Monte Carlo algorithm for symmetric stochastic matrices (in which case $\norm{A}_1 = 1$). For problems such as the simulation of diffusive discrete-time Markov chains, where $\norm{A^t u}^{-1} = O(\sqrt{N})$ at large $t$, this algorithm provides a \edit{exponential} speedup in the size of matrix $A$ over Ref.~\cite{apers2018quantum}. The second algorithm has a run-time of $\tilde{O}(Dt^2\mathrm{poly}\left(\norm{v}\norm{u}\varepsilon^{-1}\right))$  to compute $v^\dagger A^\tau u$ for all $\tau \in \{0, 1, 2 \dots t\}$. While this is slower than the quantum-walk based algorithm, it only uses Hamiltonian simulation as a primitive and thus is more suitable for near-term quantum hardware. It also achieves a run-time comparable to the quantum algorithms based on linear equation solve \cite{Berry2014ODE, berry2017quantum}. Furthermore, for matrices that are not stochastic and consequently cannot be classically powered with the Monte Carlo algorithm, this algorithm achieves a quantum speedup over repeated matrix multiplication since its run-time does not scale polynomially with the size of the matrix. Finally, following a construction similar to Ref.~\cite{berry2007efficient}, we provide no-go theorems that limit the speedups achievable with a quantum computer for powering non-Hermitian matrices.

\section{Problem definition, prelimnaries and summary of results}\label{sec:prelim}
We consider the problem of powering a Hermitian matrix $A \in \mathbb{C}^{N\times N}$ that is stable i.e. all of its eigenvalues have a magnitude less than 1, \edit{or equivalently $\norm{A}_2 \leq 1$}. Furthermore, we will assume the matrix to be $D-$sparse i.e. every row or column of the stochastic matrix has at most $D$ non-zero elements. Hermitian matrices arising in practice will typically have $D = O(1)$ or $O(\text{polylog}(N))$. The matrix powering problem that we consider is precisely defined below.\\ \ \\
\noindent\textbf{Problem (Matrix powering)} \emph{Given a $D-$sparse \edit{stable} Hermitian matrix $A \in \mathbb{C}^{N \times N}$, a positive integer power $t$ and vectors $v, u \in \mathbb{C}^N$, compute $v^\dagger A^t u$ to a specified precision $\varepsilon > 0$.} \\

\noindent We point out that previous works that solve the matrix powering problem in various contexts adopt a different problem definition wherein they aim to prepare a quantum state encoding $A^t u$. Since in many application of matrix powering we are finally interesting in computing its inner product, $v^\dagger A^t u$, with another vector $v$ which is typically known beforehand, the algorithms proposed in this paper directly compute this expectation value without ever explicitly prepare a quantum state encoding $A^t u$. We make two further notes about this problem definition:
\begin{enumerate}
    \item The precision of the output of this algorithm is assumed to be in a probabilistic sense i.e.~the algorithm is said to produce an estimate $X$ of a quantity $x$ with precision $\varepsilon$ if $\text{Prob}[|X - x| \leq \varepsilon]$ is large enough. The value of this probability, often referred to as the \emph{confidence level} of the algorithm, is assumed to be a pre-specified constant close to 1 throughout this paper and we will suppress it in the complexity results.
    \item We assume a black-box query model for the sparse Hermitian matrix $A$ i.e.~we assume access to two oracles $O_F$ and $O_A$ which allow us to access the elements of the Hermitian matrix. The oracle $O_F$ provides access to the indices of the non-zero elements of each column of the Hermitian matrix via the implementation of a unitary that satisfies
    \begin{align}
        O_F\ket{j, k} = \ket{j, f(j, k)} \ \forall j\in[N], k \in [D],
    \end{align}
    where $f(j, k)$ is the index of the $k^\text{th}$ non-zero element in the $j^\text{th}$ row or column. The oracle $O_A$ provides access to the non-zero elements of the matrix $A$ via the implementation of a unitary that satisfies
    \begin{align}
        O_A \ket{j, k}\ket{z} = \ket{j, k}\ket{z \oplus A_{j, k}} \ \forall j, k \in [N],
    \end{align}
    \edit{where $A_{j, k}$ are the complex elements of the matrix $A$ that are represented by a bit-string upto some specified precision $\delta$. On a quantum computer, these oracles can be implemented with quantum circuits of depth $O(D\ \text{polylog}(1 / \delta))$ \cite{berry2009black}. On near-term hardware, there might be alternative more efficient ways of implementing these oracles for specific matrices $A$ (for instance the matrices corresponding to local Hamiltonians of a lattice of classical spins). In this paper, for clarity, we will express our complexity results in terms of the number of calls to the oracles $O_F$ and $O_A$ and these can easily be translated to the circuit depths for various hardware-specific implementations.}
\end{enumerate}

In the remainder of this paper, we provide several quantum algorithms to solve the matrix powering problem and achieve speedups over classical algorithms. The first algorithm builds on Ref.~\cite{apers2018quantum} and combines a quantum walk together with a Hadamard test and a classical sampling algorithm to obtain the following result.
\begin{theorem}
\label{theorem:classical_sampling}
Given a constant $C>0$ such that $\norm{A}_1 < C$, the matrix-powering problem can be solved with a quantum algorithm in $O(C^{2t} D\sqrt{t}\left\Vert v\right\Vert^{2}\left\Vert u\right\Vert ^{2}\varepsilon^{-2})$
calls to the oracles $O_F, O_A$.
\end{theorem}
Furthermore, by employing the linear combination of unitaries (LCU) technique along with quantum amplitude amplification \cite{Montanaro2015Montecarlo}, we can obtain a quadratic improvement in the scaling of the run-time with the precision $\varepsilon$.
\begin{theorem}
\label{theorem:LCU_projective}
Given a constant $C>0$ such that $\norm{A}_1 < C$, the matrix-powering problem can be solved with a quantum algorithm in $\tilde{O}(C^{t} D\sqrt{t}\left\Vert v\right\Vert \left\Vert u\right\Vert \varepsilon^{-1})$
calls to the oracles $O_F, O_A$.
\end{theorem}
Here $\tilde{O}$ hides any polylog complexity factors. \edit{We point out that these algorithms suffer from an exponential scaling with the power $t$ when $\norm{A}_1 > 1$ --- this is due to the fact that the quantum walk construction we employ can only be used if the sum of magnitude of the elements of each row (or column) of $A$ is smaller than 1. For a number of matrix powering problems, such as simulation of discrete-time Markov chains, $\norm{A}_1 = 1$, and the run-time of these scales sublinearly with $t$.} The above results improve the fast-forwarding algorithm presented in Refs.~\cite{apers2018quantum}, whose run-time scales inversely with $\norm{A^t u}$ --- our approach avoids this scaling at the expense of scaling with $\norm{u}^2, \norm{v}^2$. This could be of relevance in problems such as the simulation of diffusive discrete-time Markov chains, where $\norm{A^t u}^{-1} = O(\sqrt{N})$ at large $t$. Furthermore, compared to the quantum algorithms based on linear-equation solve, this result has a quadratic speedup in $t$. We also remark that when compared to classical algorithms, we obtain an exponential speedup in $N$ over matrix multiplication algorithm and a quadratic speedup in $t$ over the Monte Carlo algorithm when the matrix $A$ is stochastic.

While the quantum walk based algorithms provided above are able to achieve `fast-forwarding', i.e.~a sublinear run-time with respect to the matrix power $t$, they are difficult to implement on near-term quantum hardware. Given experimental constraints, it is widely believed that Hamiltonian simulation \cite{Lloyd1996simulators,Berry2015hamiltonian,Childs2012HamiltonianLCU} will be one of the first problems to be solved on practical hardware.  Furthermore, simulation of several classes of Hamiltonians can also be implemented on analog quantum simulators \cite{georgescu2014quantum, aspuru2012photonic, buluta2009quantum} which are significantly easier to experimentally build as compared to fully-programmable quantum computers. Based on a truncated Fourier series expansion of the function $f(x) = x^t$, we provide a quantum algorithm to solve the matrix-powering problem with only the ability to use Hamiltonian simulation.
\begin{theorem}
\label{theorem:hamiltonian_simulation}
The matrix powering problem can be solved simultaneously for all powers from $0$ to $t$  using an efficient Hamiltonian simulator in time  $\tilde{O}\left(t^2\mathrm{poly}\left(\norm{v}\norm{u}\varepsilon^{-1}\right)D\right)$.
\end{theorem}
We note that this result has a worse run-time not only when compared to the quantum-walk algorithms, but also with classical Monte Carlo algorithm if the matrix $A$ is stochastic. However, it achieves an exponential speedup in $N$ over classical repeated matrix multiplication algorithm although at an expense of quadratically worse scaling with $t$, for matrices that are not stochastic. Furthermore, it achieves the same run-time as quantum algorithms based on the linear-equation solve if they are employed to compute matrix powers from $0$ to $t$. The key advantage of this algorithm over other quantum algorithms is its feasibility to being implemented on near-term quantum hardware.

Finally, all the algorithms provided above assume the matrix $A$ to be Hermitian, in which case it was possible to obtain a `fast-forwarding' speedup using quantum walks i.e.~compute $A^t$ in time $\Theta(\sqrt{t})$. A natural question to ask is if fast-forwarding is possible for non-Hermitian matrices as well. By utilizing a construction similar to the no-go theorems for Hamiltonian simulation \cite{berry2007efficient} and relying on the result that even a quantum computer cannot speedup the calculation of parity of $N-$bits \cite{farhi1998limit, beals2001quantum}, we provide the following no-go theorem.
\begin{theorem} [No-go theorem]
\label{theorem:no_go_theorem_irr}
There cannot exist a quantum algorithm that solves the matrix-powering problem in\\ $\tilde{O}\big(t^\alpha\emph{poly}(\norm{u}, \norm{v}, \varepsilon^{-1})\big)$ calls to the oracle $O_F, O_A$, with $\alpha < 1$, for any arbitrary irreducible sparse matrix $A$.
\end{theorem}
We point out that while these no-go theorems rigorously show that it is not possible to fast-forward the matrix-powering problem for generic non-Hermitian matrices, it does not prohibit an improvement of the run-time's dependence on the size of the matrix. Indeed, matrix-powering methods based on quantum linear equation solvers \cite{Berry2014ODE, berry2017quantum} obtain an exponential improvement over classical  algorithms even for non-Hermitian matrices if the matrix is not stochastic.

The remainder of this paper contains proofs of the theorems stated above. In section \ref{sec:fast-forwarding} we describe the matrix-powering algorithms presented in this paper and prove theorems \ref{theorem:classical_sampling}, \ref{theorem:LCU_projective} and \ref{theorem:hamiltonian_simulation}. In section \ref{sec:no-go-theorems}, we prove the no-go theorem \ref{theorem:no_go_theorem_irr}. We only provide proofs of the most important theorems in the main text, and details are relegated to the appendices.

\section{Matrix multiplication algorithm}\label{sec:fast-forwarding}

Before detailing the matrix-powering algorithm, we provide the following lemma that map the computation of $v^\dagger A^t u$ to the overlap of $A^t$, $\bra{\psi}A^t\ket{\psi}$, with quantum states $\ket{\psi}$ that depend on $u, v$. This transformation is useful since the Hadamard test naturally allows for the computation of such overlaps.
\begin{lemma}
\label{lemma:lemma_exp_eval}
Given a Hermitian matrix $A$ and vectors $v, u$, it follows that
\begin{align*}
    &\emph{Re}[v^\dagger A^t u] = \frac{1}{2}\bigg(\lambda_1^R \bra{\psi_1^R} A^t \ket{\psi_1^R} + \lambda_2^R \bra{\psi_2^R} A^t \ket{\psi_2^R}\bigg),\\
     &\emph{Im}[v^\dagger A^t u] = \frac{1}{2}\bigg(\lambda_1^I \bra{\psi_1^I} A^t \ket{\psi_1^I} + \lambda_2^I \bra{\psi_2^I} A^t \ket{\psi_2^I}\bigg),
\end{align*}
where $\ket{\psi_{i}^R}$ and $\lambda_i^R$, $i \in \{1, 2\}$, are the eigenvectors and eigenvalues of the Hermitian matrix $uv^\dagger + v u^\dagger$ and $\ket{\psi_{i}^I}$ and $\lambda_i^I$, $i \in \{1, 2\}$, are the eigenvectors and eigenvalues of the Hermitian matrix $i(v u^\dagger - uv^\dagger)$.
\end{lemma}
\textbf{Proof}: It follows immediately from the Hermiticity of $A$ that $2\text{Re}[v^\dagger A^t u] = \text{Tr}[A^t(u v^\dagger + v u^\dagger)]$ and $2\text{Im}[v^\dagger A^t u] = \text{Tr}[iA^t(v u^\dagger - u v^\dagger)]$. Using $u v^\dagger + v u^\dagger = \lambda_1^R \ket{\psi_1^R}\bra{\psi_1^R} + \lambda_2^R \ket{\psi_2^R}\bra{\psi_2^R}$ and $i(vu^\dagger - uv^\dagger) = \lambda_1^I \ket{\psi_1^I}\bra{\psi_1^I} + \lambda_2^I \ket{\psi_2^I}\bra{\psi_2^I}$, we obtain the result in the lemma.

Consequently, we will focus on developing methods to compute the overlap $\bra{\psi}A^t\ket{\psi}$ efficiently. It can also be noted that for problems where $u$ and $v$ are sparse, the eigenvectors $\ket{\psi_{1, 2}}$ introduced above are also sparse and consequently efficiently preparable on quantum computers. Furthermore, we note that the eigenvalues $\lambda_1$ and $\lambda_2$ are bounded by the norms $u$ and $v$, which is concretely stated in the following lemma.
\begin{lemma}
\label{lemma:eig_vals_bounds}
All eigenvalues $\lambda$ of $uv^\dagger + vu^\dagger$ and $i(vu^\dagger - uv^\dagger)$ satisfy \edit{$|\lambda| \leq 2\norm{u} \norm{v}$}.
\end{lemma}
\textbf{Proof}: Denoting by $\ket{\psi}$ the normalized eigenvector corresponding to the eigenvalue $\lambda$, it follows that $|\lambda| = \norm{ \big(u v^\dagger + v u^\dagger\big) \ket{\psi} } \leq \norm{u} \ |v^\dagger\ket{\psi}| + \norm{v}\  |u^\dagger\ket{\psi}| \leq 2 \norm{u} \norm{v}$. A similar proof holds for the eigenvalues of $i(vu^\dagger - uv^\dagger)$.
\subsection{Fast-forwarding with quantum walks}\label{subsec:qwalk}

One of the key ingredients in the quantum walk based algorithms for the matrix powering problem is expressing $A^t$ as a linear combination of Chebyshev polynomials of $A$,
\begin{align}\label{eq:chebyschev_poly_exp}
    A^t =\sum_{m=0}^{t}p_{m}T_{m}\left(A\right),
\end{align}
where $p_m$ is a probability distribution given by
\begin{align}\label{eq:binomial_distr}
    p_m = \begin{cases}
    \frac{1}{2^{t - 1}} {t \choose (t - m) / 2} & \text{for }m>0, t = m \text{ mod }2\\
    \frac{1}{2^t} {t \choose t / 2} & \text{for } m = 0, t = 0 \text{ mod }2 \\
    0 & \text{otherwise}
    \end{cases}.
\end{align}

Consequently, a quantum circuit to compute the overlap $\bra{\psi}A^t\ket{\psi}$ can be constructed from a quantum circuit that can compute the overlap $\bra{\psi}T_m(A)\ket{\psi}$ for a specified $m \in \{0, 1 \dots t\}$. As is shown below, this can be done with a quantum walk provided that the 1-norm of $A$ is smaller than $1$. Since this isn't necessary for stable Hermitian matrices, we assume that we have access to an upper bound $C$ on this norm i.e. $\norm{A}_1 \leq C$ and compute $\bra{\psi}(A/C)^t\ket{\psi}$, \edit{albeit to a precision $C^t$ higher than that required in $\bra{\psi}A^t\ket{\psi}$}. Therefore, in the remainder of this section, unless otherwise mentioned, we will assume $\norm{A}_1 \leq 1$.

A quantum walk construction similar to that used in Refs.~\cite{apers2018quantum, szegedy2004qwalk, watrous2001qwalk} together with a Hadamard test allows us to compute these overlaps. However, since the elements of the matrix $A$ can be complex, it is important to design the quantum walk with care so as to account for the phase of the complex matrix elements \cite{berry2009black}.  For $A \in \mathbb{C}^{N \times N}$, we consider a Hilbert space $\mathbb{C}^{N+1} \otimes \mathbb{C}^{N+1}\otimes\mathbb{C}^2$ and assume access to a unitary $V$ that satisfies
\begin{widetext}
\begin{subequations}
\begin{align}
   &V\ket{i, 0, 0} =\sum_{k=1}^{N}\sqrt{|A_{k,i}|}e^{i\varphi_{k, i}/2}\left|i,k,1\right\rangle +\bigg(1-\sum_{k=1}^{N}\left|A_{k,i}\right|\bigg)^{1/2}\left|i,N+1,1\right\rangle  \ \text{if } i\neq N+1,\\
    &V^{\dagger}\ket{i, j, 1} = \sqrt{|A_{j,i}|}e^{-i\varphi_{j, i} / 2}\left|i,0,0\right\rangle +\left|\phi^{\perp}\right\rangle \left|1\right\rangle \ \text{for some }\ket{\phi^\perp}\text{ if } i \neq N+1,\\
    &V\ket{N+1,j,b}=\ket{N+1,j,b}.
 \end{align}
\end{subequations}
\end{widetext}
where if $A_{i, j} = |A_{i, j}|e^{i\angle A_{i, j}}$ for $\angle A_{i, j} \in (-\pi, \pi] $ then $\varphi_{i, j} = \angle A_{i, j}$ for $i \geq j$ and $-\angle A_{i, j}$ for $i< j$. Furthermore, we introduce the operator $S$ given by
\begin{align}
    S\ket{i, j, b} =\begin{cases} \ket{i, j, 0} \ \text{if } b = 0,\\
    \ket{j, i, 1} \ \text{if } b = 1 \ \text{and } i \neq j, \\
    \mathrm{sign}\left(A_{i,i}\right)\ket{i, i, 1} \ \text{if } b = 1 \ \text{and } i=j.
    \end{cases}.
\end{align}
We remark that this operator is different from that used in Ref.~\cite{apers2018quantum} --- in particular, we have modified this operator to account for possibly negative on-diagonal elements of the matrix $A$ which Ref.~\cite{apers2018quantum} did not handle since they were dealing with a stochastic matrix. Finally, the quantum walk operator $W$ can then be constructed using the operators $V, S$ and a reflection about the last qubit
\begin{align}\label{eq:qwalk_operator}
W = -(I \otimes I \otimes \sigma_z)V^\dagger S V.
\end{align}
We then obtain the following lemma, similar to that obtained in Refs.~\cite{apers2018quantum, apers2019quantum} stating that $m$ applications of the quantum walk operator effectively applies $T_m(A)$ on an input state conditioned on the state of the last qubit. 

\begin{lemma}[Quantum walk for $T_m(A)$]
\label{lemma:qwalk_tm}
The unitary operator $W$ defined in Eq.~\ref{eq:qwalk_operator} satisfies
\begin{align}
    W^m \ket{\psi}\ket{0}\ket{0} = T_m(A)\ket{\psi}\ket{0}\ket{0} + \ket{\psi^\perp}\ket{1}
\end{align}
for some $\ket{\psi^\perp} \in \mathbb{C}^{N+1} \otimes \mathbb{C}^{N+1}$.
\end{lemma}
As is shown in appendix \ref{app:oracle_mappings}, the quantum walk operator $A$ can be implemented with $O(D)$ calls to the oracles $O_F, O_A$ that access the matrix $A$. In order to estimate the overlap $\bra{\psi}T_m(A)\ket{\psi}$ using the walk operator $W$, we introduce a controlled version of $W$, $W^c$ via
\begin{align}
    W^c = I \otimes \ket{0}\bra{0} + W \otimes \ket{1}\bra{1}.
\end{align}
We then have the following lemma to compute the overlap $\bra{\psi}T_m(A)\ket{\psi}$ using a hadamard test with the controlled operator $W^c$.
\begin{lemma}[Chebyshev polynomial overlap]
\label{lemma:qwalk_hadamard}
Consider the state $(W^c)^m \ket{\psi}\ket{0, 0, +}$, and measure the last two qubits on the basis $\{\ket{0, +}, \ket{0, -}, \ket{1, +}, \ket{1, -}\}$. Define a random variable $X_m$ based on the measurement outcome $\mu$ via
\[
X_m = \begin{cases}
+1 & \text{ if } \mu = (0, +) \\
-1 & \text{ if } \mu = (0, -) \\
0 & \text{ otherwise }
\end{cases}
\]
then $E(X_m) = \bra{\psi} T_m(A) \ket{\psi}$.
\end{lemma}
While this overlap estimation procedure can be used together with the Chebyshev polynomial expansion in Eq.~\ref{eq:chebyschev_poly_exp} to compute $\bra{\psi}A^t\ket{\psi}$, this would not provide any fast-forwarding since computing $T_t(A)$ would require $t$ applications of the operator $W^c$. However, an important insight into the nature of the coefficients $p_m$ in the expansion in Eq.~\ref{eq:binomial_distr} is that they concentrate around $m \sim \sqrt{t}$. One possible approach to exploit this property is to sample $m$ from the probability distribution given by the coefficients $p_m$ in Eq.~\ref{eq:binomial_distr}, and compute the overlap $\bra{\psi}T_m(A)\ket{\psi}$ of the corresponding Chebyshev polynomial using lemma \ref{lemma:qwalk_hadamard} --- this would allow us to reduce, on an average, the number of times the walk operator $W^c$ is applied. This is formalized in the lemma below:

\begin{lemma}[Matrix power overlap with classical sampling]
\label{lemma:smat_olap}
The overlap $\bra{\psi}A^t\ket{\psi}$ can be computed by estimating the mean of a random variable $X$ which is generated by first sampling $m \in \{0, 1 \dots t\}$ from the probability distribution in Eq.~\ref{eq:binomial_distr}, followed by drawing a sample of $X_m$ defined in lemma \ref{lemma:qwalk_hadamard} using the state $\ket{\psi}$. Furthermore, this estimation can be done with a precision $\varepsilon$ with $O(\varepsilon^{-2}\sqrt{t})$ calls to the controlled walk operator $W^c$ or equivalently with $O(D\varepsilon^{-2}\sqrt{t})$ calls to the oracles $O_F, O_A$.
\end{lemma}
\textbf{Proof}:
We can immediately see that
\begin{align}
    E(X) &= \sum_{m=0}^t p_m E(X_m) \nonumber\\
         &= \sum_{m=0}^t p_m \bra{\psi}T_m(M)\ket{\psi} =\bra{\psi}M^t \ket{\psi} 
\end{align}
Furthermore, noting that $X^2 \in \{0, 1\}$, it follows that $\text{var}(X) \leq 1$. Consequently, $E(X)$ can be estimated to a precision of $\varepsilon$ with $N = O(1 / \varepsilon^2)$ samples. Furthermore, the average number of calls to the controlled walk operator $W^c$, $\langle m \rangle$ is given by
\begin{align}
    \langle m \rangle = \sum_{m = 0}^t m p_m = \frac{1}{2^{t}} \sum_{n = 0}^{t/2} n {t \choose t/2-n}=\frac{2+t}{2^{t-2}} {t \choose t/2-1}.
\end{align}
For large $t$, using Stirling's approximation this is
\begin{align}
    {t \choose t/2-1}=\frac{t/2}{t/2+1}{t \choose t/2}\sim{t \choose t/2}\sim\frac{2^t}{\sqrt{t\pi /2}}.
\end{align}
Therefore $\left\langle m\right\rangle =O\left(\sqrt{t}\right)$, and consequently the number of calls to $W^c$ operator to achieve a precision $\varepsilon$ given by $N \langle m\rangle = O(\varepsilon^{-2}\sqrt{t})$.\\

\noindent\textbf{Proof of theorem 1:} Combining lemma \ref{lemma:smat_olap} with lemma \ref{lemma:lemma_exp_eval}, we obtain a procedure for solving the matrix powering problem. The complexity result in theorem \ref{theorem:classical_sampling} can be obtained as follows: Given an upper bound $C$ on $\norm{A}_1$, we note that to compute $v^\dagger A^t u$ to precision $\varepsilon$, we need to compute $\bra{\psi_{1, 2}^{L, R}}(A/C)^t \ket{\psi_{1, 2}^{L, R}}$ to a  precision at-most $\varepsilon / 4 \norm{u}\norm{v} C^t$ which can be done using the algorithm in lemma \ref{lemma:lemma_exp_eval} with $O(D\varepsilon^{-2}\sqrt{t} \norm{v}^2 \norm{u}^2 C^{2t})$ calls to the oracles $O_F, O_A$.

While the algorithm described above allows a quadratic fast-forwarding for the matrix powering problem, the dependence of the run-time on the precision $\varepsilon$ can also be improved by using amplitude amplification, which is precisely stated in the following lemma from Ref.~\cite{Montanaro2015Montecarlo}:

\begin{lemma}[Overlap estimation, Theorem 2.5 of Ref.~\cite{Montanaro2015Montecarlo}] 
\label{lemma:amplitude_amp}
Given a state $\ket{\psi}$ in terms of its preparation unitary $U$ from a known state $\ket{0}$: $\ket{\psi} = U\ket{0}$, an observable $V$ and an estimate $\sigma$ of its variance satisfying $ \bra{\psi}V^2\ket{\psi} - (\bra{\psi}V\ket{\psi})^2 \leq \sigma^2$, $\bra{\psi} V\ket{\psi}$ can be estimated on a quantum computer with precision $\varepsilon$ with $\tilde{O}(\varepsilon^{-1}\sigma )$ calls to the unitary $U$.
\end{lemma}
In order to use amplitude amplification and achieve fast forwarding in the same algorithm, we approximate the problem of computing the overlap $\bra{\psi}A^t\ket{\psi}$ for a given $\ket{\psi}$ to computing an overlap of the form $\bra{\phi_t}V\ket{\phi_t}$ where the operator $V$ is independent of $t$ and the state $\ket{\phi_t}$ can be prepared in $\Theta(\sqrt{t})$ calls to the oracles $O_F, O_A$. This is achieved by using a combination of quantum walks with the hadamard test and the linear combination of unitaries (LCU) technique similar to Ref.~\cite{apers2018quantum}. The quadratic fast-forwarding in this approach is also obtained due to the concentration of the coefficients $p_m$ in Eq.~\ref{eq:binomial_distr} around $m = \sqrt{t}$. This is made concrete by the following lemma, according to which the sum in Eq.~\ref{eq:chebyschev_poly_exp} can be truncated after $\sim O(\sqrt{t}\ \text{log}(\varepsilon^{-1}))$ terms while incurring a specified additive error $\varepsilon$.

\begin{lemma}
\label{lemma:truncation_matrixnorm}
If $A$ is a stable Hermitian matrix and $\ket{\psi}$ is a normalized state, then $\forall \varepsilon > 0$ and $C\geq2\log({2}/{\varepsilon})$
\[
\left|\left\langle \psi\right|A^{t}\left|\psi\right\rangle -\left\langle \psi\right|\sum_{m=0}^{\sqrt{Ct}}p_{m}T_{m}\left(A\right)\left|\psi\right\rangle \right|\leq\varepsilon
\]
\end{lemma}
\textbf{Proof}: Using lemma 3 from Ref.~\cite{apers2018quantum}, we obtain that $\forall \varepsilon > 0, C \geq 2\log(2 / \varepsilon)$
\begin{align}\label{eq:chebyschev_trunc_ineq}
   \bigg | x^t - \sum_{m = 0}^{\sqrt{Ct}}p_m T_m(x) \bigg | \leq \varepsilon \ \forall x \in [-1, 1].
\end{align}
Since $A$ is a stable matrix, its eigenvalues will have a magnitude less than equal to 1. Furthermore, since $A$ is also Hermitian, its eigenvalues are real and lie in $[-1, 1]$ and hence satisfy Eq.~\ref{eq:chebyschev_trunc_ineq}. Denoting by $\lambda_i, \ket{\phi_i}$ the eigenvalues and eigenvectors of $M$ and using its spectral decomposition $A = \sum_{i}\lambda_i \ket{\phi_i}\bra{\phi_i}$ we obtain
\begin{align}
    \bigg|\bra{\psi}A^t \ket{\psi}& - \bra{\psi} \sum_{m=0}^{\sqrt{Ct}}p_{m}T_{m}\left(A\right) \ket{\psi}\bigg|\leq \nonumber \\
    & \sum_{k} |\bra{\phi_k}\psi\rangle|^2\bigg|\lambda_k^t - \sum_{m=0}^{\sqrt{Ct}} p_m T_m(\lambda_k)\bigg|  \leq \varepsilon
\end{align}
Consequently, we can effectively approximate $A^t$ as weighted linear combination of $O(\sqrt{t})$ Chebyshev polynomials of $A$ --- while the quantum walk operator introduced in Eq.~\ref{eq:qwalk_operator} can be used to individually implement the Chebyshev polynomials, in order to implement their linear combination we use the LCU technique \cite{Childs2012HamiltonianLCU}. Below, we show the construction of an operator to effectively apply $\sum_{m = 0}^{\tau} p_m W^m$ to given quantum state. We do this by introducing auxillary qubits and implementing the following unitary $V_P$ depending on the coefficients $p_m$:
\begin{align}\label{eq:VP}
    &V_P\ket{\phi}\ket{0, 0} =\nonumber\\
    &\sum_{m = 0}^\tau \sqrt{p_m}\ket{\phi}\ket{m, 0} + \bigg[1 - \sum_{m = 0}^\tau p_m\bigg]^{1/2} \ket{\phi}\ket{0, 1}
\end{align}
Furthermore, we assume access to a controlled quantum walk operator $W_\tau$ defined by
\begin{align}\label{eq:wtau_lcu}
    W_\tau = \sum_{m = 0}^\tau W^m \otimes \ket{m}\bra{m} \otimes I
\end{align}
The operator $W_\tau$ requires $\tau$ calls to the quantum walk operator $W$. Therefore, following the result in appendix \ref{app:oracle_mappings}, it can be constructed with $O(D\tau)$ calls to the oracles $O_F, O_A$. The operator $V_P^\dagger W_\tau V_P$ then effectively applies the linear combination $\sum_{m=0}^\tau p_\tau W^\tau$ to an input state.
\begin{lemma}[LCU adapted from Refs.~\cite{apers2018quantum, Childs2012HamiltonianLCU}]
\label{lemma:LCU}
The unitary operator $V_P^\dagger W_\tau V_P$, with $V_P$ and $W_\tau$ defined in Eqs.~\ref{eq:VP} and \ref{eq:wtau_lcu} respectively, satisfy 
\begin{align}
    V_P^\dagger W_\tau V_P \ket{\phi}\ket{0, 0} = \sum_{m = 0}^\tau p_m W^m \ket{\phi}\ket{0, 0} + \ket{\phi^\perp}\ket{1},
\end{align}
for some $\ket{\phi_\perp}$.
\end{lemma}
In order to compute $\left\langle \psi\right|\sum_{m=0}^{\tau}p_{m}T_{m}\left(A\right)\left|\psi\right\rangle $, we consider a controlled version of the operator defined in lemma \ref{lemma:LCU}: $W_\tau^{c}=I\otimes\left|0\right\rangle \left\langle 0\right|+V_P^\dagger W_\tau V_P\otimes\left|1\right\rangle \left\langle 1\right|$. It then follows from lemmas \ref{lemma:qwalk_tm} and \ref{lemma:LCU} that computing the expectation value of the operator $\ket{0}\bra{0}\otimes \sigma_z$ on the last two qubits in the circuit of $W_\tau^c$ on a state prepared by application of $H W_\tau^c H$ (where $H$ is a hadamard gate on the last qubit) to $\ket{\psi} \ket{0, 0 \dots 0}$ allows us to evaluate $\left\langle \psi\right|\sum_{m=0}^{\tau}p_{m}T_{m}\left(A\right)\left|\psi\right\rangle $. By truncating the linear combination to an appropriate number of terms using lemma \ref{lemma:truncation_matrixnorm} and using amplitude estimation (lemma \ref{lemma:amplitude_amp}), we obtain the following lemma for estimating the overlap of the matrix with a given state.
\begin{lemma}[Matrix power overlap with LCU] For a Hermitian stable matrix $A$ with $\norm{A}_1 \leq 1$ and a state $\ket{\psi}$, $\bra{\psi} A^t \ket{\psi}$ can be computed on a quantum computer to a precision $\varepsilon$ with $\tilde{O}(D \sqrt{t} \varepsilon^{-1}) $ calls to the oracles $O_F, O_A$.
\label{lemma:smat_olap_lcu}
\end{lemma}
\textbf{Proof}: The overlap estimation can be done using the Hadamard test described above --- as shown in lemma \ref{lemma:truncation_matrixnorm}, it is sufficient to use $\tau = O(\sqrt{t}\ \text{log}( \varepsilon^{-1})) $ in the LCU construction described in lemma \ref{lemma:LCU}. Furthermore, the outcome of the Hadamard test has a variance bounded by 1 and consequently a direct application of amplitude amplification (lemma \ref{lemma:amplitude_amp}) allows us to obtain an estimate of $\bra{\psi}A^t\ket{\psi}$ with $\tilde{O}(D\sqrt{t} / \varepsilon)$ calls to the oracles $O_F, O_A$.\\

\noindent\textbf{Proof of theorem 2}: Combining lemma \ref{lemma:smat_olap_lcu} with lemma 
\ref{lemma:lemma_exp_eval}, we obtain a procedure for solving the matrix powering problem.
The complexity result in theorem \ref{theorem:LCU_projective} can be obtained as follows: Given an upper bound $C$ on $\norm{A}_1$, we note that to compute $v^\dagger A^t u$ to precision $\varepsilon$, we need to compute $\bra{\psi_{1, 2}^{L, R}}(A/C)^t \ket{\psi_{1, 2}^{L, R}}$ to a  precision at-most $\varepsilon / 4 \norm{u}\norm{v} C^t$ which can be done using the algorithm in lemma \ref{lemma:lemma_exp_eval} with $\tilde{O}(\sqrt{t}\norm{v} \norm{u} C^t \varepsilon^{-1})$ calls to the oracles $O_F, O_A$.

\subsection{Matrix powering with hamiltonian simulation}\label{subsec:hamil_sim}

In this section, we describe an approach to solve the matrix powering problem using Hamiltonian simulation as a primitive and prove the complexity result in theorem \ref{theorem:hamiltonian_simulation}. Formally for our purposes, a Hamiltonian simulation can be considered to be the problem of computing $\bra{\psi}e^{-iHt}\ket{\psi}$ to a precision $\varepsilon$ given access to the sparse Hamiltonian $H$ and the states $\ket{\psi}$. A Hamiltonian simulator (implemented on a quantum computer or an analog quanutm simulator) is said to be efficient if it can solve this problem in $\tilde{O}(\text{poly}(\varepsilon^{-1}) D\norm{H}_\text{max}t)$ time, where $D$ is the sparsity of the Hamiltonian and $\norm{H}_\text{max}$ is its maximum magnitude element. State of the art algorithms for Hamiltonian simulation on quantum computers achieve such run-times for general sparse Hamiltonians, while such run-times can be achieved on quantum simulators for local Hamiltonians.

We again restrict ourselves to stable Hermitian matrices $A$. In order to compute an overlap $\bra{\psi}A^t\ket{\psi}$, we expand $A^t$ into a fourier series --- as is shown in the following two lemmas, this can be done to a precision of $\varepsilon$ while retaining only $N_h = O(t/\varepsilon)$ harmonics.
\begin{lemma}
\label{lemma:fourier_expansion}
$\forall \varepsilon \in (0, 2 / \pi)$, $N_h \geq  4t / \pi^2 \varepsilon$, $\exists a \in \mathbb{C}^{2N_h + 1}$ such that
\begin{align}\label{eq:fourier_truncation}
    \bigg|x^t - \sum_{n = -N_h}^{N_h} a_n e^{in\pi x / 2}\bigg| \leq \varepsilon \ \forall \ x\in[-1, 1],
\end{align}
where $a_n$ are the elements of the vector $a$ and $a$ can be computed classically in $O(N_ht)$ time. Furthermore, $\norm{a}_1 \leq 1 \ \forall t > 0, N_h > 0$.
\end{lemma}
A detailed proof of this lemma, as well as an explicit calculation of the coefficient vector $a$, is provided in appendix \ref{app:fourier_series}. Employing this result, we can now compute the overlap of the power $\bra{\psi}A^t\ket{\psi}$ using an efficient Hamiltonian simulator.

\begin{lemma}[Matrix overlap with Hamiltonian simulation] \label{lemma:H-simulation-expectedvalue}
Given a $D-$sparse stable Hermitian matrix $A$, $\bra{\psi}A^\tau\ket{\psi}$ can be computed for all $\tau \in \{0, 1 \dots t\}$ using an efficient Hamiltonian simulator in time $\tilde{O}(\emph{poly}( \varepsilon^{-1})Dt^2)$.
\end{lemma}
\textbf{Proof}: Since the matrix $A$ is stable and Hermitian, all of its eigenvalues are real and lie in $[-1, 1]$ thus satisfying Eq.~\ref{eq:fourier_truncation}. Denoting by $\lambda_k, \ket{\phi_k}$ the eigenvalues and eigenvectors of $A$ and using lemma \ref{lemma:fourier_expansion}, we obtain that
\begin{align}
    \bigg | &\bra{\psi}A^{\tau}\ket{\psi} -\sum_{n=-N_h}^{n=N_h}a_{n}\bra{\psi}e^{i\pi nA/2}\ket{\psi}\bigg | \nonumber\\  &\bigg |\sum_{k} |\bra{\phi_k}{\psi}\rangle|^2\bigg(\lambda_k^\tau -\sum_{n=-N_h}^{n=N_h}a_{n}e^{i\pi n\lambda_k/2}\bigg)\bigg | \leq \frac{\varepsilon}{2},
\end{align}
for an appropriately chosen $N_h = O(\tau / \varepsilon)$. Furthermore, we note that since $A$ is stable, the magnitude of all of its elements is at-most 1 i.e. $\norm{A}_\text{max} \leq 1$. Using an efficient Hamiltonian simulator, we can estimate $\bra{\psi}e^{i\pi n A / 2}\ket{\psi}$ to a precision $\varepsilon / 2$ in time $O(\text{poly}(\varepsilon^{-1})D n)$. Since $\norm{a}_1 \leq 1$, we can then compute $\sum_{n = -N_h}^{N_h} a_n \bra{\psi}e^{i\pi n A / 2}\ket{\psi}$ to a precision $\varepsilon$ on a classical computer, thus determining $\bra{\psi}A^\tau \ket{\psi}$ to a precision $\varepsilon$ --- since we need to compute $\bra{\psi}e^{i\pi n A/2}\ket{\psi}$ for $n \in \{-N_h, -N_h + 1 \dots N_h - 1, N_h\}$, the total time taken for this computation is $O(\text{poly}(\varepsilon^{-1})DN_h^2) = O(\text{poly}(\varepsilon^{-1}D)\tau^2)$. Furthermore, we note that since we propose to compute the overlaps $\bra{\psi}e^{in\pi A / 2}\ket{\psi}$ individually for all $n$, we can compute $\bra{\psi}A^\tau \ket{\psi}$ for all $\tau \in \{0, 1 \dots t\}$ with the same set of Hamiltonian simulations in time $O(\text{poly}( \varepsilon^{-1})Dt^2)$.\\

\noindent\textbf{Proof of theorem \ref{theorem:hamiltonian_simulation}}: Combining lemma \ref{lemma:H-simulation-expectedvalue} with \ref{lemma:lemma_exp_eval}, we obtain a procedure for solving the matrix powering problem. To obtain the complexity result in theorem \ref{theorem:hamiltonian_simulation}, we note that computing $v^\dagger A^\tau u$ to precision $\varepsilon$, we need to compute $\bra{\psi_{1, 2}^{L, R}}A^\tau \ket{\psi_{1, 2}^{L, R}}$ to precision at-most $\varepsilon/4\norm{u}\norm{v}$. This can be done using the algorithm in \ref{lemma:H-simulation-expectedvalue} for $\tau \in \{0, 1, 2\dots t\}$ simultaneously in  $O(\text{poly}( \varepsilon^{-1}\norm{v}\norm{u})Dt^2)$ time.

\section{No-go theorems for fast forwarding}\label{sec:no-go-theorems}
\begin{figure*}
    \centering
    \includegraphics[scale=0.4]{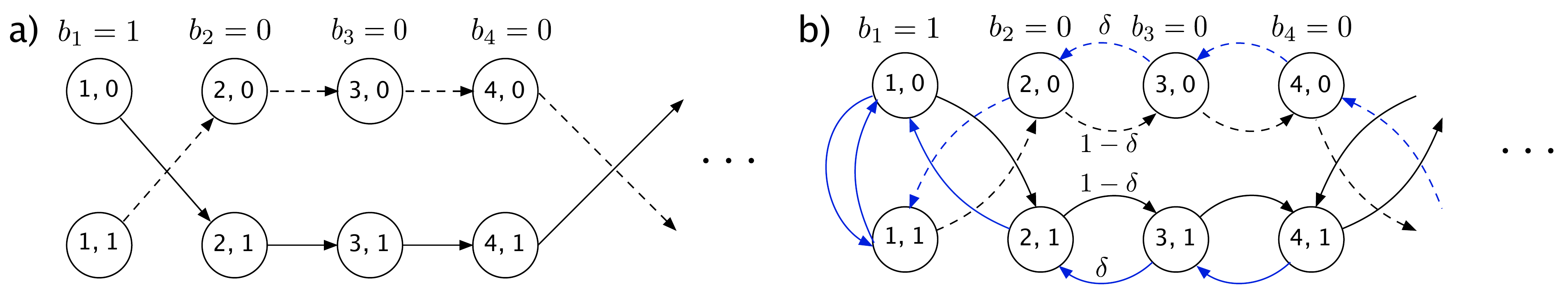}
    \caption{Schematic representation of a (a) reducible Markov chain and an (b) irreducible Markov chain that solves the $N-$bit parity problem. In (b), the blue lines indicate a stochasatic matrix element $\delta$ and the black lines indicate a stochastic matrix element $1-\delta$.}
    \label{fig:no_go_theorem}
\end{figure*}
In this section, we provide no-go theorems stated in section \ref{sec:prelim}, showing that fast forwarding the matrix powering problem is not possible for a generic non-Hermitian matrix. These no-go theorems utilize a construction similar to the no-go theorems for Hamiltonian simulation \cite{berry2007efficient}, and rely on the fact that even a quantum computer cannot speedup the calculation of parity of $N-$bits \cite{farhi1998limit, beals2001quantum}, a result concretely stated as the following Lemma:
 

\begin{lemma}[$N-$bit parity problem, Refs.~\cite{farhi1998limit, beals2001quantum}]
\label{lemma:no_go_parity}
Consider $N-$bits $b_1, b_2 \dots b_N$ which can be accessed as an oracle $O_B$: $O_B\ket{i, b} = \ket{i, b \oplus b_i}$. There cannot exist a quantum algorithm that can determine the parity $b_1 \oplus b_2 \oplus \dots b_N$ with success probability greater than $1/2$ with fewer than $N / 2$ calls to the oracle $O_B$.
\end{lemma}
We note that this result rules out even an approximate solution of the $N-$bit parity problem on a quantum computer with run-time less than $O(N)$. In particular, since the parity is known to be an integer, if an algorithm can estimate this parity to a precision $\varepsilon < 1/2$ with a confidence level greater than 1/2, then it would have solved the $N-$bit parity problem --- consequently, in the rest of this analysis we can consider the precision $\varepsilon = O(1)$. As is shown in the no-go theorem below, we can construct a matrix such that computing its $N^\text{th}$ power, in the sense specified in section \ref{sec:prelim}, allows us to solve the $N-$bit parity problem, thereby ruling out the possibility of designing a quantum algorithm that can achieve a run-time scaling sublinearly with the matrix power. We first provide a proof of lemma \ref{theorem:no_go_theorem_general}, which rules out the possibility of fast-forwarding the powering of a general matrix, and then strengthen it to obtain theorem \ref{theorem:no_go_theorem_irr} which rules out the possibility of fast-forwarding even the powering of irreducible matrices. \\ \ \\
\begin{lemma}[No-go theorem for arbitrary matrix]
\label{theorem:no_go_theorem_general}
There cannot exist a quantum algorithm that solves the matrix powering problem in $\tilde{O}\big(t^\alpha\emph{poly}(\norm{u}, \norm{v}, \varepsilon^{-1})\big)$ calls to the oracles $O_F, O_A$, with $\alpha < 1$, for any sparse matrix $A$.
\end{lemma}
\noindent\textbf{Proof}: Given $N$ bits $b_1, b_2 \dots b_N$, we can construct the following matrix powering problem that determines the parity of $b_1 \oplus b_2 \dots \oplus b_N$ on computing its $N^\text{th}$ power:
\begin{enumerate}
    \item The matrix $A\in \mathbb{C}^{2N \times 2N}$ with the rows (or columns) being indexed by $(i, \sigma)$ where $i \in [N]$ and $\sigma \in \{0, 1\}$ and matrix elements being given by
    \begin{align}\label{eq:M0_nogo}
        A_{i, \sigma; i', \sigma'} = \begin{cases}
        1 & \text{if } i' = i - 1 \ \text{and} \ \sigma \oplus \sigma' = b_{i - 1}\\
        0 & \text{otherwise}
        \end{cases}
    \end{align}
    \item The vectors $u$ and $v$ are given by
    \begin{align}\label{eq:v_u_nogo}
    &u_{i, \sigma} = \begin{cases} 1 & \text{ if } i = 0 \text{ and } \sigma = 0\\
    0 & \text{otherwise}
    \end{cases}, \nonumber \\ 
    &v_{i, \sigma} = \begin{cases}
        1 & \text{if } i = N \text{ and } \sigma = 0 \\
        -1 & \text{if } i = N \text{ and } \sigma = 1 \\
        0 & \text{ otherwise }
    \end{cases}.
\end{align} 
\end{enumerate}
The matrix $A$ is a stochastic matrix corresponding to a discrete-time Markov chain (DTMC) shown in Fig.~\ref{fig:no_go_theorem}a --- the states of this DTMC can be grouped as per their $\sigma$ index, and for every bit $b_i$ is identified with a flip of the `$\sigma$' index at $i$. Consequently, on computing $v^\dagger A^t u$, we can count the number of bits that are 1 thereby computing $b_1 \oplus b_2 \dots \oplus b_N$. Furthermore, we can note that the oracles $O_F, O_A$ can be constructed with $O(1)$ calls to the oracle $O_B$ since each bit determines the positions of the non-zero elements in at-most two columns. Furthermore, we note that $\norm{u}, \norm{v} = O(1)$ by construction. Consequently, if there existed a quantum algorithm to solve the DTMC simulation problem with $\tilde{O}(t^\alpha \text{poly}(\norm{u}, \norm{v}, 1 / \varepsilon))$ queries to $O_F, O_A$ with $\alpha < 1$, it could solve the $N-$bit parity problem in $\tilde{O}(N^\alpha)$ queries to $O_B$. This contradicts lemma \ref{lemma:no_go_parity} thus proving that no such quantum algorithm can exist.

While this argument proves that no quantum algorithm can exist to fast-forward the matrix powering problem for a general matrix, we note that the specific matrix $A$ used in this argument is reducible. Consequently, this raises the question of whether their exists a quantum algorithm that can fast-forward the powering of arbitrary irreducible matrices. We show that this too isn't possible by constructing an irreducible stochastic matrix that is very close to the reducible stochastic matrix constructed above, and thus approximately solves the $N-$bit parity problem.\\

\noindent \textbf{Proof of theorem \ref{theorem:no_go_theorem_irr}}: We consider an instance of the matrix-powering problem with $u, v$ as defined in Eq.~\ref{eq:v_u_nogo} and a matrix $A_\delta = A + B \delta$ where $A$ is defined in Eq.~\ref{eq:M0_nogo}, $\delta \in (0, 1)$ and matrix $B$ has elements given by:
\begin{align}
    B_{i, \sigma; i', \sigma'} = \begin{cases} -1 & \text{if } i' = i - 1 \text{ and } \sigma \oplus \sigma' = b_{i - 1} \\
    1 & \text{if } i' = i + 1 \text{ and } \sigma \oplus \sigma' = b_{i} \\
    1 & \text{if } i = i' \in \{1, N\}, \sigma \neq \sigma' \\
    0 & \text{otherwise }
    \end{cases}
\end{align}
The matrix $A_\delta$ is the stochastic matrix corresponding to a discrete-time Markov chain shown in Fig.~\ref{fig:no_go_theorem}b. It is easy to see that $A_\delta$ is irreducible for $\delta\neq 0$. Furthermore, we can easily bound the difference between the result of powering $A_\delta$ and $A$:
\begin{align}
    \big|v^\dagger A^t_\delta u - v^\dagger A^t u \big| &\leq \norm{v} \norm{u} \big((\norm{A}_2 + \norm{B}_2\delta)^t - \norm{A}_2^t\big)\nonumber \\ & \leq \norm{v}\norm{u}\norm{A}_2^t \big(e^{\norm{B}_2\delta t / \norm{A}_2} - 1\big).
\end{align}
For the choice of $u, v$ under consideration, $\delta = \text{log}(1 + \varepsilon / 2\sqrt{2}) / 8N^2$ and $t = N$ it follows that
\begin{align}
    |v^\dagger A_\delta^N u - v^\dagger A^N u| \leq  \frac{\varepsilon}{2}
\end{align}
wherein we have used $\norm{v} = \sqrt{2}$ and $\norm{u} = 1$ and $1 / \sqrt{2N} \leq \norm{A}_2 \leq 1$, $\norm{B}_2\leq 4\sqrt{2N}$. This shows that being able to compute $v^\dagger A_\delta^t u$ to precision $\varepsilon / 2$  allows us to determine $v^\dagger A^t u = b_1 \oplus b_2\oplus \dots \oplus b_N$ to precision $\varepsilon$. Consequently, from lemma \ref{theorem:no_go_theorem_general}, the no-go theorem follows for irreducible matrices as well.

\section{Conclusion}
This paper studied the problem of computing the power of a stable Hermitian matrix. Following the construction of Ref.~\cite{apers2018quantum}, we show that fast-forwarding is possible while powering stable Hermitian matrices and present algorithms based on quantum walks that improve their results. We also present a complementary algorithm to solve the matrix powering problem using only Hamiltonian simulators which could potentially be used on near-term quantum hardware. Finally, by establishing a map between the  the $N-$bit parity determination problem to a matrix powering problem, we show that quantum computers cannot fast-forward powering of non-Hermitian matrices.

\begin{acknowledgements}
We  acknowledge  support  from the  ERC  Advanced  Grant  QUENOCOBA  under  the EU  Horizon  2020  program  (grant  agreement  742102) and  from  the  Deutsche  Forschungsgemeinschaft  (DFG, German Research Foundation) under the project number414325145 in the framework of the Austrian Science Fund(FWF): SFB F7104.

\end{acknowledgements}
\appendix
\section{Costs for quantum walk operators in terms of matrix oracles}\label{app:oracle_mappings}

In this appendix we show how the quantum walk operator $W$ can be implemented with $O(D)$ calls to the oracles $O_F$, $O_A$ that access the matrix $A$, where $D$ is the sparsity. The quantum walk operator is $W=V^{\dagger}SV$. We have assumed access to a unitary operator $V$ such that
\begin{widetext}
\begin{subequations} 
\begin{align} \label{eq:operator_V}
   &V\ket{i, 0, 0} =\sum_{k=1}^{N}\sqrt{|A_{k,i}|}e^{i\varphi_{k, i}/2}\left|i,k,1\right\rangle +\bigg(1-\sum_{k=1}^{N}\left|A_{k,i}\right|\bigg)^{1/2}\left|i,N+1,1\right\rangle  \ \text{if } i\neq N+1,\\
    &V^{\dagger}\ket{i, j, 1} = \sqrt{|A_{j,i}|}e^{-i\varphi_{j, i} / 2}\left|i,0,0\right\rangle +\left|\phi^{\perp}\right\rangle \left|1\right\rangle \ \text{for some }\ket{\phi^\perp}\text{ if } i \neq N+1,\\
    &V\ket{N+1,j,b}=\ket{N+1,j,b}.
 \end{align}
\end{subequations}
\end{widetext}
We can write $V$ as
\begin{align}
    V=\ket{N+1}\bra{N+1}\otimes I\otimes I+\sum_{i=1}^{N}\left|i\right\rangle \left\langle i\right|\otimes V_{i}
\end{align}
Therefore, $V$ applies $V_i$ depending on the second and third registers depending on the state of the first qubit. The $V_i$ act as
\begin{align}
    V_{i}=\left|\psi_{i}\right\rangle \left\langle 0,0\right|+\sum_{j=1}^{N}\left|j,0\right\rangle \left\langle j,0\right|+\sum_{j=1}^{N+1}\left|\psi_{i,j}^{\perp}\right\rangle \left\langle j,1\right|
\end{align}

 $\left|\psi_{i,j}^{\perp}\right\rangle $, along with $\left|\psi_{i}\right\rangle $, form an orthonormal basis. The expression for $\ket{\psi_i}$ is
 
\begin{align}
    \ket{i}\left|\psi_{i}\right\rangle &=U\left|i,0,0\right\rangle \nonumber \\ &=\sum_{k=1}^{N}\sqrt{A_{k,i}^*}\left|i,k,1\right\rangle +\sqrt{1-\sum_{k=1}^{N}\left|A_{k,i}\right|}\left|i,N+1,1\right\rangle 
\end{align}

 where the square root is defined as in \ref{eq:operator_V}. Therefore, we need to show that we can apply the operator $U$ that prepares $\ket{\psi_i}$ with $O(D)$ queries to the oracles. This can be proved with a slight modification of the procedure in \cite{Chiang2010qwalkcircuits}.
 
 To do this, we start with  the state $\ket{i}\ket{0} $. We then prepare a list with all the neighbours of $i$: $\left|i\right\rangle \left|f\left(i,1\right)\right\rangle ...\left|f\left(i,D\right)\right\rangle $. This can be done with $O(D)$ calls to $O_F$. Now we can prepare a list containing all the non-zero probabilities in the i-th row $\left|i\right\rangle \left|A_{i,f\left(i,1\right)}\right\rangle ...\left|A_{i,f\left(i,D\right)}\right\rangle $. This can be done with $O(D)$ calls to $O_M$. Appending an extra register to this list, we can compute $\ket{i}\ket{A_{i,f(i,1)}}\ket{A_{i,f(i,D)}}\ket{1-\sum_{k=1}^{D}\left|A_{i,f\left(i,k\right)}\right|}$
 Using both lists, following the procedure in \cite{Chiang2010qwalkcircuits}, we can prepare the state
 
 \begin{align}
     \sum_{j=1}^{D}\sqrt{A_{i,f\left(i,j\right)}^*}\left|i\right\rangle \left|j\right\rangle + \sqrt{1-\sum_{k=1}^{D}\left|A_{i,f\left(i,k\right)}\right|} \ket{i,N+1}
 \end{align}
Querying again $O_F$ we obtain the desired state. Therefore, it follows that the quantum walk operator $W$ can be implemented with $O(D)$ calls to $O_F$ and $O_A$.

\section{Proof of lemma \ref{lemma:fourier_expansion}}\label{app:fourier_series}
\noindent In this appendix, we provide a proof of lemma \ref{lemma:fourier_expansion} introduced and used in the main text. Our goal is to calculate a fourier series expansion of $x^t$ which converges pointwise when $x \in [-1, 1]$, and estimate the number of terms of the fourier series that we need to retain to achieve a certain precision in this expansion. We point out that for odd $t$, a fourier series expansion of $x^t$ in the interval $[-1, 1]$ could exhibit a Gibb's phenomena at $x = \pm 1$ since the periodic extension of $x^t$ is  not continuous. Consequently, we instead consider the function $f(x)$ on the interval $[-2, 2]$ defined below
\begin{align}
    f_t(x) = \begin{cases}
            x^t & \text{for } |x| \leq 1, \\
            (2 - x)^t & \text{for } 1 < x \leq 2, \\
            (-2 - x)^t & \text{for } -2 \leq x < -1.
    \end{cases}
\end{align}
The periodic extension of this function is continuous, since $f_t(2) = f_t(-2) \ \forall t$ and $f_t(x)$ is continuous within the interval $(-2, 2)$. Furthermore, for $|x| \leq 1$ this function coincides with $x^t$. We can now write down a fourier series expansion for this function which converges point-wise for all $x \in [-2, 2]$ --- for ease of analysis, we treat the cases when $t$ is even and odd seperately:
\begin{align}\label{eq:fourier_seried_even_odd_def}
    f_t(x) = 
    \begin{cases}
    \sum_{p = 0}^\infty c_p(t)\cos(p \pi x) & \text{ if } t \text{ is even}, \\
    \sum_{p = 0}^\infty s_p(t) \sin((2p + 1)\pi x / 2) & \text{ if } t \text{ is odd},
    \end{cases}
\end{align}
where
\begin{subequations}\label{eq:fourier_coefficients_def}
\begin{align}
    c_p(t) &= \int_0^2 f_t(x) \cos(p\pi x) dx \nonumber \\ &= 2\int_0^1 x^t \cos(p \pi x) dx,\\
    s_p(t) &= \int_0^2 f_t(x) \sin((2p + 1) \pi x / 2) dx \nonumber \\ &= 2\int_0^1 x^t \sin((2p + 1) \pi x / 2) dx.
\end{align}
\end{subequations}
We point out that the Eq.~\ref{eq:fourier_seried_even_odd_def} can be rewritten in terms of complex exponentials to obtain a fourier series of the form used in lemma \ref{lemma:fourier_expansion}. Furthermore, $c_p(t)$ and $s_p(t)$ can be explicitly evaluated to obtain
\begin{subequations}\label{eq:fourier_coefficients}
\begin{align}
    &c_p(t) = 2(-1)^p\sum_{k=1}^{t / 2}\frac{(-1)^{k + 1}}{(p^2 \pi^2)^k} \prod_{i = 0}^{2k - 2} (t - i)\\
    &s_p(t) =2(-1)^p\sum_{k = 1}^{(t + 1) / 2} \frac{(-1)^{k + 1}}{((p + 1/2)^2 \pi^2)^k}\prod_{i=0}^{2k - 2}(t - i)
\end{align}
\end{subequations}
We point out that $s_p(t)$ and $c_p(t)$ can be computed in $O(t)$ time on a classical computer using a recursive implementation of the summations in Eq.~\ref{eq:fourier_coefficients}. We now consider a truncated fourier series expansion i.e.~we construct the function $\hat{f}_t^N(x)$ from the coefficients $c_p(t), s_p(t)$ where
\begin{align}
        \hat{f}^N_t(x) = 
    \begin{cases}
    \sum_{p = 0}^N c_p(t)\cos(p \pi x) & \text{ if } t \text{ is even}, \\
    \sum_{p = 0}^N s_p(t) \sin((2p + 1)\pi x / 2) & \text{ if } t \text{ is odd}.
    \end{cases}
\end{align}
We then obtain that $\forall x \in [-1, 1]$, $|x^t - \hat{f}_t^N(x)| \leq e_N$, where
\begin{align}
    e_N(t) =  \begin{cases}
    \sum_{p = N + 1}^\infty |c_p(t)| & \text{ if } t \text{ is even},\\
    \sum_{p = N + 1}^\infty |s_p(t)| & \text{ if } t \text{ is odd}.
    \end{cases}
\end{align}
It now remains to provide bounds on $e_N(t)$ in terms of $t$ and $N$. We note from Eq.~\ref{eq:fourier_coefficients} that
\begin{align}
    \forall p > t/ \pi, \ |c_p(t)| &\leq 2 \sum_{k = 1}^{t / 2} \frac{1}{(p^2 \pi^2)^k} \prod_{i = 0}^{2k - 2}(t - i) \nonumber\\ &\leq 2\sum_{k = 1}^{\infty} \frac{t^{2k - 1}}{(p^2\pi^2)^k} = \frac{2t}{p^2 \pi^2 - t^2}.
\end{align}
A similar bound holds for $|s_p(t)|$:
\begin{align}
     \forall p > t/ \pi, \ |s_p(t)| &\leq 2\sum_{k=1}^{(t + 1)/2}\frac{1}{((p + 1/2)^2\pi^2)^k}\prod_{i = 0}^{2k - 2}(t - i) \nonumber \\ &\leq 2\sum_{k = 1}^{\infty}\frac{t^{2k - 1}}{(p^2\pi^2)^k} \leq \frac{2t}{p^2 \pi^2 - t^2}
\end{align}
Consequently, it then follows that
\begin{align}\label{eq:error_estimate_even}
    \forall N > t / \pi, \ e_N(t) &\leq 2\sum_{p = N + 1}^\infty \frac{t}{p^2 \pi^2 - t^2} \nonumber\\ &\leq \int_{N}^\infty \frac{2t}{x^2 \pi^2 - t^2} dx = \frac{1}{\pi } \log\bigg(\frac{N \pi + t}{N \pi - t}\bigg).
\end{align}
To ensure that $e_N(t)$ is smaller than a given precision $\varepsilon$, we can then choose $N$ to be
\begin{align}\label{eq:N_limit}
    N = \frac{t}{\pi \tanh(\pi\varepsilon / 2)} \geq \frac{2t}{\pi^2\varepsilon}.
\end{align}
We point out that for this estimate to be correct, the chosen $N$ should also be larger than $t / \pi$ (Eq.~\ref{eq:error_estimate_even}), which is implied by Eq.~\ref{eq:N_limit} if the precision $\varepsilon$ to be smaller than $2 / \pi$ and we obtain the estimate provided in lemma \ref{lemma:fourier_expansion}.

Finally, we compute the $l_1$ norm of the coefficients $c_p(t)$ and $s_p(t)$. We note that $(-1)^p c_p(t) \geq 0$ and $(-1)^ps_p(t)\geq 0$ for all $p \geq 0$. This is easily seen as follows --- from Eq.~\ref{eq:fourier_coefficients_def}, using integration by parts it follows that
\begin{subequations}
\begin{align}\label{eq:recursion_fourier_coeff}
    &c_p(t) = 2(-1)^p\frac{t}{p^2 \pi^2} - \frac{t(t-1)}{p^2 \pi^2} c_{p}(t-2)\  \nonumber \\   &s_p(t) = 2(-1)^p \frac{t}{(p + 1 /2)^2 \pi^2} - \frac{t(t-1)}{(p + 1 /2)^2 \pi^2}s_p(t - 2).
\end{align}
\end{subequations}
Furthermore, from Eq.~\ref{eq:fourier_coefficients_def}, it also follows that
\begin{align}
    |c_p(t)|, |s_p(t)| \leq 2 \int_0^1 x^t dx = \frac{2}{t + 1} \ \forall t\geq 0,
\end{align}
from which it follows that $2t \geq |t(t-1)c_p(t - 2) |$ and $2t \geq |t(t - 1)s_p(t - 2)|$. Together with Eq.~\ref{eq:recursion_fourier_coeff}, it follows that $(-1)^p c_p(t) \geq 0$ and $(-1)^p s_p(t)\geq 0$ for all $p \geq 0$. Therefore,
\begin{align*}
    &\sum_{p = 0}^\infty |c_p(t)| = \sum_{p=0}^\infty c_p(t) (-1)^p = \sum_{p=0}^\infty c_p(t) \cos p\pi = 1 \text{ and }\\
    &\sum_{p=0}^\infty |s_p(t)| = \sum_{p=0}^\infty s_p(t) (-1)^p = \sum_{p=0}^\infty s_p(t) \sin\bigg(\frac{(2p + 1)\pi}{2}\bigg) = 1.
\end{align*}
Therefore, the 1-norm of $[c_0(t), c_1(t) \dots c_N(t)]$ and $[s_0(t), s_1(t) \dots s_N(t)]$ is less than 1.
\bibliography{references.bib}
\end{document}